\documentstyle[prd,aps,epsfig,preprint]{revtex}
\tighten
\begin{document}
\thispagestyle{empty}
\baselineskip24pt
\draft
\begin{center}
{\large \bf 
Hara Theorem, GIM Model, and Asymmetry in Weak Radiative 
Hyperon Decays}
\end{center}
 \begin{center}
{Elena~N.~Bukina$^a$, Vladimir~M.~Dubovik$^a$ and
Valery~S.~Zamiralov$^{b}$}
\hbox{\it $^a$Joint Institute for Nuclear Research,
          141980 Dubna, Moscow region, Russia} 
\hbox{\it $^b$ D.V. Skobeltsyn Institute of Nuclear Physics,
Moscow State University, Moscow, Russia}
\end{center}
\date{\today}
\begin{abstract}
It is shown that in the framework of GIM model,
the parity-violating (PV)
$\Sigma^{+}(uus) \Rightarrow p(uud) + \gamma $ decay  amplitude
does not vanish but just
transforms into another PV decay amplitude, namely, of the
$\Omega^{+}_{cc}(ccs)\rightarrow \Xi^{+}_{cc}(ccd)+\gamma $ decay.
The known result is used that 
the relevant part of the CP-invariant effective current$\times$ current
Hamiltonian changes sign under simultaneous quark changes
$d\leftrightarrow s$ and $u\leftrightarrow c$. 
So, asymmetry of $\Sigma^{+} \Rightarrow p + \gamma $ decay   
should not vanish in contrast to the Hara result of the
old-fashioned $SU(3)_{f}$ model. 
At the same time, the zero asymmetry prediction persists for the
$ \Xi^{-} \Rightarrow \Sigma^{-} + \gamma $ decay.
\end{abstract}

\section*{Introduction}

$\quad$ Weak radiative decays of hyperons
were first analyzed theoretically
forty years ago \cite{Behr}-\cite{Furl}. 
In 1964, in the framework of the unitary symmetry model,
a theorem was proved by Hara that
decay asymmetry in charged hyperon weak radiative decays 
$\Sigma^{+} \Rightarrow p + \gamma $ 
and $ \Xi^{-} \Rightarrow \Sigma^{-} + \gamma $ should  vanish 
in the limit of exact $SU(3)_{f}$
\cite{Hara}.
Experimental discovery of a large negative asymmetry 
in the radiative decay $ \Sigma^{+} \Rightarrow p + \gamma $  \cite{Ger}, 
confirmed later \cite{PDG}, stimulated efforts 
to explain a drastic contradiction between  experimental results and 
the Hara zero asymmetry prediction
( see, e.g., \cite{ZenM} and references therein).
The problem is actual up to now as shown by appearance of new approaches
to cite among the most recent ones \cite{Barry}.

Another problem is related to inconsistency between $ SU(3)_{f}$ symmetry
\cite{Hara} and quark model predictions \cite{Ryaz}
for the asymmetry value in hyperon weak 
radiative decays.
Quark models, while more or less succeeding in describing experimental
data on branching ratios and asymmetry parameters (see, e.g., a review 
\cite{ZenM} and references therein),
did not reproduce the Hara claim 
in the $SU(3)_{f}$ symmetry limit.

The origin of this discrepancy is under discussion up to now although many
authors investigated this problem thoroughly 
\cite{Ryaz}-\cite{Dmitra96} 
(see also \cite{ZenM} for a very complete list of publications).
Recently, it 
was a subject of discussion in \cite{Zendm} and \cite{Dmitra99}.
Both sides with fine arguments proved once more that the 
inconsistency persists though the authors have different points of view as
to the origin of it. In  \cite{Dmitra96} and \cite{Dmitra99}, 
it was argued that while the covariant
amplitude in \cite{Ryaz} was taken gauge-invariant, the 
nonrelativistic reduction performed there proved to be gauge-variant. 
Instead, in \cite{Zendm}, it was stated
that gauge invariance was preserved in  \cite{Ryaz}, and the origin 
of discrepancy should be hidden in other rather implicit assumptions as
that of a sufficiently localized current. 

On the other hand we showed that the problem
may be solved in part by envoking the toroid dipole moment \cite{BDZ},
although hardly it is possible to attain  experimentally observed value.

We would like to show that the origin of both the problems
can be in the specific formulation of basic assumptions 
on weak interaction taken in \cite{Hara}
in the framework of the $SU(3)_{f}$ symmetry model, and
a remedy to the problems can be found already by taking 
a four-flavor model.

\section {The basic Hara result}

To describe strangeness-changing weak radiative
hyperon decays of the baryon octet $B\rightarrow B^{\prime}+\gamma,$
in \cite{Hara},  usual basic assumptions as to the 
character of weak interactions were used :

(1) The effective weak interaction Hamiltonian is 
a sum of products of charged weak currents and their 
Hermitian conjugates.

(2) The weak interaction is CP-invariant.
(Known violation of the CP-invariance was ignored .)

These assumptions were read in \cite{Hara} as invariance
of the effective strangeness-changing weak interaction Hamiltonian 
under exchange of the unitary indices $2\leftrightarrow 3$
(see also \cite{Marsh}):
$$
H^{eff}_{SU(3)_{f}}(|\Delta S|=1)=
\frac{G_{F}}{2\sqrt{2}}sin \theta_{C} cos \theta_{C}
\left\{[J_{1}^{2},J_{3}^{1}]_{+}+(2\leftrightarrow 3)\right\},
$$
where $ J_{1}^{2}$ and $J_{1}^{3}$ are weak hadron currents with
$|\Delta S|=0$ and $|\Delta S|=1$, respectively, in the usual
$SU(3)_{f}$ tensor notation.
Let us rewrite it in terms of the weak quark currents.
The effective CP-invariant quark weak Hamiltonian with $|\Delta S|=1$ 
in the three-quark model is
\begin{equation}
H^{eff}_{W}(|\Delta S|=1)=\frac{G_{F}}{\sqrt{2}}sin \theta_{C} cos \theta_{C}
\left\{(\bar u O_{\mu}d)(\bar s O_{\mu}u)+
(\bar u O_{\mu}s)(\bar d O_{\mu}u)\right\},
\end{equation}
where $O_{\mu}=\gamma_{\mu}(1-\gamma_{5}).$
The $H^{eff}_{W}(|\Delta S|=1)$ is invariant under 
exchange of s- and d-quarks,
$s\leftrightarrow d$ (or equivalently, under exchange 
of indices  $2\leftrightarrow 3$ 
because of the relation 
$
B^{\alpha}_{\beta}=
\epsilon_{\beta \gamma \delta}\{q^{\beta},q^{\gamma}\}q^{\delta},
$
where $\alpha, \beta, \gamma, \delta=1,2,3$ and 
$u=q^{1},d=q^{2},s=q^{3}$,
between the baryon wave functions and quark ones.)

The parity-violating (PV) parts of the amplitudes of weak
radiative hyperon decays were written 
in \cite{Hara} as follows: 
\begin{eqnarray}
M^{PV}_{SU(3)_{f}}=J^{(d)}_{\mu} \epsilon_{\mu} + H.C.= \nonumber \\
\left\{a^{d}(\overline{B}^{3}_{2}O^{d}_{\mu}B^{1}_{1}-
\overline{B}^{1}_{1}O^{d}_{\mu}B^{2}_{3})+ 
b^{d}(\overline{B}^{3}_{1}O^{d}_{\mu}B^{1}_{2}-
\overline{B}^{2}_{1}
O^{d}_{\mu}B^{1}_{3})+
\right. \label{hara}\\
c^{d}(\overline{B}^{1}_{2}O^{d}_{\mu}B^{3}_{1}-\overline{B}^{1}_{3}
O^{d}_{\mu}B^{2}_{1})+
\left.
d^{d}(\overline{B}^{1}_{1}O^{d}_{\mu}B^{3}_{2}-
\overline{B}^{2}_{3}O^{d}_{\mu}B^{1}_{1})\right\}\epsilon_{\mu},
\nonumber
\end{eqnarray}
with the gauge-invariant Lorentz structure $ O^{d}_{\mu}=
i \sigma_{\mu \nu} k_{\nu}\gamma_{5} $ \cite{Hara} ,
$\epsilon_{\mu}$ being the photon polarization 4-vector.
The upper script $d$ stays for {\it dipole transition moment}.
To preserve invariance under exchange $2\leftrightarrow 3$, 
Hara put $a^{d}=-d^{d}$,
opening a possibility of the nonzero asymmetries for
$(\Sigma^{0},\Lambda)\rightarrow n+\gamma $ and
$\Xi^{0}\rightarrow (\Sigma^{0},\Lambda)+\gamma $ decays.
At the quark level, a requirement of the CP-invariance in the form
$d\leftrightarrow s$ just prescribes that 
amplitudes of the decays 
$(\Sigma^{0},\Lambda)(usd)\rightarrow n(ddu)+\gamma $ (+its HC)
transform into the HC amplitudes of quite different decays
$\Xi^{0}(ssu)\rightarrow (\Sigma^{0},\Lambda)(uds)+\gamma. $

Instead of this, PV transition amplitudes of the
decays $ \Sigma^{+} \rightarrow p+\gamma$ and $ \Xi^{-}\rightarrow\Sigma^{-}
+\gamma $ disappear in Eq.(\ref{hara}). The relevant terms 
$ \overline{B}^{3}_{1}O^{d}_{\mu}B^{1}_{2}$ and $\overline{B}^{1}_{2}
O^{d}_{\mu}B^{3}_{1},$ which should be invariant under exchange of indices
$2\leftrightarrow 3$, change signs under Hermitian conjugation; therefore,
the condition $b^{d}=c^{d}=0 $ should be imposed.
The origin of this result is readily seen at the quark level.
Under the exchange $d\leftrightarrow s$, 
the PV amplitudes of
decays $ \Sigma^{+}(uus) \rightarrow p(uud)+\gamma$ 
and $ \Xi^{-}(ssd)\rightarrow\Sigma^{-}(dds)
+\gamma $ are transformed into the respective HC amplitudes of the
same decays but with a wrong sign. 

This is in fact a source of all the troubles with 
the hyperon weak radiative decays in the framework of the $SU(3)_{f}$
model, as it just gives zero asymmetry for            
$ \Sigma^{+} \rightarrow p+\gamma$ and $ \Xi^{-}\rightarrow\Sigma^{-}
+\gamma $ decays.

\section {GIM model and PV amplitudes of the weak radiative hyperon decays}

The simple picture of the weak interaction used in \cite{Hara} 
was in fact based on a single weak isodublet  
$\left(\begin{array}{cc}u\\d_{C} 
\end{array}\right)_{L}$,
and led, as it is well known, to the existence of the strangeness-
changing neutral weak current. This difficulty unknown in 1964
was overcome in the famous GIM model \cite{GIM} where another
weak isodublet 
$\left(\begin{array}{cc}c\\s_{C} \end{array}\right)_{L}$,
was introduced in order to make the neutral weak current
diagonal in quark flavors. Following this we try to rewrite
the Hara theorem in the context of the four-flavor scheme
( for a moment not taking a six-flavor one) in order to get
rid of the undesirable strangeness-changing neutral current
implicitly hidden in \cite{Hara}.

In the framework of the
GIM model, the relevant part of the CP-invariant effective
Hamiltonian containing $|\Delta S|=1$ piece reads  now in terms of quarks as
\begin{eqnarray}
H^{eff}_{GIM}=
\frac{G_{F}}{\sqrt{2}}sin \theta_{C} cos \theta_{C}
\left\{(\bar u O_{\mu}d+\bar c O_{\mu}s)(\bar s O_{\mu}u-\bar d O_{\mu}c)+
\right.\nonumber\\ \left.
(\bar u O_{\mu}s-\bar c O_{\mu}d)(\bar d O_{\mu}u+\bar s O_{\mu}c)\right\}.
\label{gim}
\end{eqnarray}
But the Hamiltonian $H^{eff}_{GIM}$ is no longer invariant under either
change $s\leftrightarrow d$ or $c\leftrightarrow u$. Instead of this
under a simultaneous change  $s\leftrightarrow d$ and 
$c\leftrightarrow u$, it just changes an overall  sign.
We try now to insert this 
Ansatz into the effective flavor-changing electromagnetic current.

And this turns to be a solution of the Hara puzzle.

Really, now for the $ \Sigma^{+} \rightarrow p+\gamma$ PV
decay amplitude ,
the Hermitian conjugation and invariance of 
$H^{eff}_{GIM} $ given by Eq.(\ref{gim}) under  the flavor exchange
are uncorrelated.

To see this in terms of $SU(4)_{f}$ baryon wave functions,
along the lines of the Hara proof \cite{Hara} 
and a discussion in \cite{Marsh}, we construct
an appropriate baryon $SU(4)_{f}$ weak radiative transition current 
that 

(i) conserves built-in weak interaction properties
(1) and (2); 

(ii) mantains gauge invariance .

The corresponding matrix element 
should be invariant under an
overall change of sign with simultaneous changes of indices
$1\leftrightarrow 4$ and $2\leftrightarrow 3.$ 
This statement is essentially similar to that made by Hara,
\cite{Hara} 
and independent
of the conjecture of the U- or P-spin used in \cite{Vasa}.
In this way,
(i) would be satisfied at the level
of the four-flavor model, and instead of Eq.(\ref{hara}),
we obtain:
\begin{eqnarray}
M_{GIM}^{PV}(|\Delta S|=1)=J^{PV}_{\mu} \epsilon_{\mu} + H.C.=
\nonumber\\
\left\{a^{pv}(\overline{B}^{34}_{2}O^{d}_{\mu}B^{1}_{14}-
\overline{B}^{14}_{1}O^{d}_{\mu}B^{2}_{34}-
\overline{B}^{21}_{3}O^{d}_{\mu}B^{4}_{41}+\overline{B}^{41}_{4}
O^{d}_{\mu}B^{3}_{21})+\right.\nonumber\\
b^{pv}(\overline{B}^{34}_{1}O^{d}_{\mu}B^{1}_{24}-\overline{B}^{24}_{1}
O^{d}_{\mu}B^{1}_{34}-
\overline{B}^{21}_{4}O^{d}_{\mu}B^{4}_{31}+
\overline{B}^{31}_{4}O^{d}_{\mu}B^{4}_{21})+\nonumber\\
c^{pv}(\overline{B}^{14}_{2}O^{d}_{\mu}B^{3}_{14}-
\overline{B}^{14}_{3}O^{d}_{\mu}B^{2}_{14})
+\nonumber\\
d^{pv}(\overline{B}^{14}_{1}O^{d}_{\mu}B^{3}_{24}
-\overline{B}^{24}_{3}O^{d}_{\mu}B^{1}_{14}-
\overline{B}^{14}_{4}O^{d}_{\mu}B^{2}_{13}+
\overline{B}^{13}_{2}O^{d}_{\mu}B^{4}_{14})+\nonumber\\
\left.
f^{pv}(\overline{B}^{13}_{1}O^{d}_{\mu}B^{1}_{12}
-\overline{B}^{12}_{1}O^{d}_{\mu}B^{1}_{13}+
\overline{B}^{34}_{4}O^{d}_{\mu}B^{4}_{24}-
\overline{B}^{24}_{4}O^{d}_{\mu}B^{4}_{34})
\right\}\epsilon_{\mu},
\label{newhara}
\end{eqnarray}
where $SU(4)$ 20-plet baryon wave functions can be written 
in terms of quark wave functions as usual
$$
B^{\alpha}_{\beta \gamma}=\epsilon_{\beta \gamma \eta \rho}
\{q^{\alpha},q^{\eta} \}q^{\rho},\quad \alpha,\beta,
\gamma,\eta,\rho=1,2,3,4
$$
and $u=q^{1},d=q^{2},s=q^{3},c=q^{4}.$
We remind that $SU(3)$ octet baryons are
$$B^{1}_{34}=p,\quad B^{1}_{24}=\Sigma^{+},\quad
B^{3}_{14}=\Xi^{-}, \quad B^{3}_{24}=\Xi^{0},$$
$$B^{2}_{14}=\Sigma^{-},\quad
B^{1}_{14}=\frac{1}{\sqrt{2}}\Sigma^{0}+\frac{1}{\sqrt{6}}\Lambda^{0},
\quad B^{2}_{34}=n,$$ 
while those relevant charmed ones are
$$B^{4}_{13}=\Xi^{+}_{cc}(ccd), \quad B^{4}_{12}=\Omega^{+}_{cc}(ccs), $$
$$
B^{2}_{13}=\Sigma^{0}_{c}(cdd), \quad
B^{4}_{14}=\frac{2}{\sqrt{6}}\Xi^{\prime 0}_{cs}(csd), \quad
B^{3}_{12}=\Omega^{0}_{c}(css)$$
$$ 
B^{4}_{34}=\frac{2}{\sqrt{6}}\Lambda^{+}_{c}(cud), \quad
B^{1}_{13}=\frac{1}{\sqrt{2}}\Sigma^{+}_{c}(cud)+
\frac{1}{\sqrt{6}}\Lambda^{+}_{c}(cud),$$
$$ B^{4}_{24}=-\frac{2}{\sqrt{6}}\Xi^{+}_{cs}(csu),\quad
B^{1}_{12}=\frac{1}{\sqrt{2}}\Xi^{\prime +}_{cs}(csu)+
\frac{1}{\sqrt{6}}\Xi^{+}_{cs}(csu).$$
The main point is that now no coefficient in Eq.(\ref{newhara})
should be put equal to zero, so
neither $\Sigma^{+}\rightarrow p+\gamma$
nor $\Xi^{-}\rightarrow \Sigma^{-}+\gamma$ decay
PV amplitudes vanish.

We repeat once more that, in the Hara formulation, the term  
$\overline{B}^{3}_{1}O^{d}_{\mu}B^{1}_{2}$ in Eq.(\ref{hara})
(which describes PV part of the $\Sigma^{+}\rightarrow p+\gamma$ decay )
has, as its Hermitian conjugation term (HC),
$-\overline{B}^{2}_{1}O^{d}_{\mu}B^{1}_{3}$, while if the assumption (i)
reads as invariance under change $2\leftrightarrow 3$, there must be
$+\overline{B}^{2}_{1}O^{d}_{\mu}B^{1}_{3}$. This results in a zero
contribution to the PV amplitude of the
$\Sigma^{+}\rightarrow p+\gamma$ decay . The same reasoning is valid also
for the term describing $\Xi^{-}\rightarrow\Sigma^{-}+\gamma$ decay. 

But now with the requirement (i) formulated in the form 
consistent with the GIM model, the term 
$(\overline{B}^{34}_{1}O^{d}_{\mu}B^{1}_{24}- 
\overline{B}^{24}_{1}O^{d}_{\mu}B^{1}_{34})$ corresponding
to the $\Sigma^{+}(uus)\rightarrow p(uud)+\gamma$ decay (+ its HC )
just transforms into the
term $(\overline{B}^{31}_{4}O^{d}_{\mu}B^{4}_{21}-
\overline{B}^{21}_{4}O^{d}_{\mu}B^{4}_{31})$
that describes
the PV part of the 
$\Omega^{+}_{cc}(ccs)\rightarrow \Xi^{+}_{cc}(ccd)+\gamma$ decay
amplitude (+ its HC). 

Instead of this the term
$(\overline{B}^{14}_{2}O^{d}_{\mu}B^{3}_{14}
-\overline{B}^{14}_{3}O^{d}_{\mu}B^{2}_{14})
$ describing the PV part
of the  $\Xi^{-}(ssd)\rightarrow\Sigma^{-}(dds)+\gamma$ decay 
(+ its HC) transforms into itself with the same sign. It is easy to see that
the corresponding PC part of this amplitude vanishes, so a zero asymmetry
prediction persists for this decay. But this result does not
contradict quark model calculations
where only the $s\rightarrow d+\gamma$  decay
diagram contributes, and its PV part vanishes in the limit $m_{s}=m_{d}$
\cite{Vasa,Gad}.

So, none of the PV amplitudes of the hyperon octet radiative decays
should be zero, if general requirements like properties under
Hermitian conjugation, CP-invariance, etc, are applied in the 
framework of the GIM model 
( not talking of the six-flavor scheme for a moment). 
Disregarding new baryons containing c-quark in Eq.(\ref{newhara}) 
and omitting index 4 altogether, we formally arrive at Eq.(\ref{hara}).
Also, $|\Delta S|=1$ neutral weak current appears and with it
inadmissible values for $B\rightarrow B^{\prime}+e^{+}e^{-}$ rates.
The unique mode to go to the $SU(3)$ symmetry model
limit, without falling in troubles with weak neutral currents,
would be to put the Cabibbo angle equal to zero, forbidding thus
all the processes with $|\Delta S|=1$ in the sector of quarks u,d,s.

\section {Conclusion}

So, the Hara prediction proves to be valid only in the old-fashioned
$SU(3)_{f}$ model where by default there are also strangeness-changing
neutral currents, about which almost nobody worried in 1964.
Already in the framework of the GIM model, even not taking
into account more quark flavors,
the Hara prediction of the zero asymmetry in the
$\Sigma^{+}\rightarrow p+\gamma$ decay
is no longer valid. It is interesting that 
due to vanishing of the parity-conserving amplitude,
the zero asymmetry prediction
persists for the $\Xi^{-}\rightarrow\Sigma^{-}+\gamma$ decay.

Thus, there is no contradiction of the unitary model approach
with the quark model one either for the
$\Sigma^{+}\rightarrow p+\gamma$ decay or
for the $\Xi^{-}\rightarrow\Sigma^{-}+\gamma$ one.

We conclude our letter with a remark that
maybe experimental observation of the
nonzero asymmetry in the $\Sigma^{+}\rightarrow p+\gamma$ decay 
\cite{Ger} could serve as indication of some serious difficulties 
in the description of electroweak interactions in the framework
of the 3-quark model already in early seventies and could 
be just one more argument in favor of the 4th quark and
diagonal flavor structure of weak neutral currents.

\section*{ACKNOWLEDGEMENTS}
The authors are grateful to V.~Dmitra\u{s}inovi\'c for the reprint of his 
paper.
One of the authors (V.Z.) thanks G.Costa, F.Hussain, N.Paver, S.Petcov ,
and M.Tonin for interest 
in the early stage of this work and discussion.
One of the authors (V.Z.) is grateful to the International Centre for
Theoretical physics, Trieste, where part of this work was done, for
hospitality and financial support.


\begin{references}
\bibitem{Behr} R.E.~Behrends, Phys.Rev. {\bf 111}, 1691 (1958).
\bibitem{Furl} G.~Calucci, G.~Furlan, Il Nuovo Cimento {\bf 21}, 679 (1961).
\bibitem{Hara} Y.~Hara, Phys.Rev.Lett. { \bf 12}, 378 (1964).
\bibitem{Ger} L.K.~Gershwin et al., Phys.Rev. { \bf 188}, 2077 (1969).
\bibitem{PDG} Particle Data Group,  The European Phys. J.C { \bf 3},613(1998);
Phys.Rev. { \bf D54}, 1-I (1996).
\bibitem{ZenM} J.~Lach and P.~\.{Z}enczykowski, Int.J.Mod.Phys.A {\bf 10},
              3817 (1995).
\bibitem{Barry} B.~Borasoy and B.R.~Holstein, 
Phys.Rev. {\bf D59}, 054019 (1999).
\bibitem{Ryaz} A.N.~Kamal and Riazuddin, Phys.Rev. {\bf D28}, 2317 (1983).
\bibitem{Sharma} R.C.~Verma and A.~Sharma, Phys.Rev. {\bf D38}, 1443 (1988).
\bibitem{Zen} P.~\.{Z}enczykowski, Phys.Rev. {\bf D44}, 1485 (1991);
ibid., {\bf D40}, 2290 (1989).
\bibitem{Dmitra96} V.~Dmitra\u{s}inovi\'c, 
Phys.Rev. {\bf D54}, 5899 (1996).
\bibitem{Zendm} P.~\.{Z}enczykowski, Phys.Rev. {\bf D60}, 018901 (1999).
\bibitem{Dmitra99} V.~Dmitra\u{s}inovi\'c, 
Phys.Rev. {\bf D60}, 018902 (1999).
\bibitem{BDZ} E.N.~Bukina, V.M.~Dubovik, V.S.~Zamiralov,
              Phys.Lett. {\bf B449}, 93 (1999). 
\bibitem{Marsh} R.E.~Marshak, Riazuddin and C.P.~Ryan,
Theory of Weak Interactions in Particle Physics,
Wiley-Interscience, 1969. 
\bibitem{GIM} S.~Glashow, J.~Iliopoulos, L.~Maiani, 
Phys.Rev. {\bf D2}, 1285 (1970).
\bibitem{Vasa} N.~Vasanti, Phys.Rev. {\bf D13}, 1889 (1976).
\bibitem{Gad} N.G.~Deshpande and G.~Eilam, Phys.Rev. {\bf D26}, 2463 (1982).
\end{references}
\end{document}